\documentclass[aps,pra,superscriptaddress,twocolumn]{revtex4-2}
\usepackage{bm}
\usepackage{amsmath}
\usepackage{amsthm}
\usepackage{amsfonts}
\usepackage{amssymb}
\usepackage{latexsym}
\usepackage[utf8]{inputenc}
\allowdisplaybreaks

\usepackage{slashed}
\usepackage{float}
\usepackage{dcolumn}
\usepackage{graphicx}

\newcolumntype{.}{D{x}{}{-1}}

\newcolumntype{w}[1]{D{.}{.}{#1}}

\begin{document}

		\title{Antiprotonic atoms with nonperturbative inclusion of vacuum polarization and finite nuclear mass}

	\author{Vojt\v{e}ch Patk\'o\v{s}}
	\affiliation{Faculty of Mathematics and Physics, Charles University,  Ke Karlovu 3, 121 16 Prague
		2, Czech Republic}

	\author{Krzysztof Pachucki}
	\affiliation{Faculty of Physics, University of Warsaw,
		Pasteura 5, 02-093 Warsaw, Poland}

	\begin{abstract}
		We demonstrate that energy levels of excited states in a hydrogenic system consisting of an arbitrary nucleus and an antiproton can be calculated 
		within the framework of nonrelativistic quantum electrodynamics, even for a large nuclear charge $Z$.
		It is because for rotational states the expansion parameter is $Z\,\alpha/n$.
		The main advantage of this approach is the possibility of exact inclusion of the finite nuclear mass, 
		which we achieve up to the $(Z\,\alpha)^6$ order. In addition, we  include unperturbatively the one-loop and two-loop electron vacuum polarization (evp) potentials in 
		the nonrelativistic Hamiltonian, as well as in the leading relativistic correction. 
		The obtained results for $l>1$ states of antiprotonic atoms with spinless nucleus are the most accurate to date. 
		We make available a user-friendly {\sl Mathematica} code for antiprotonic atoms {\sl PbarSpectr}, which can be further improved 
		by combining evp potentials with $(Z\,\alpha)^5$ QED effects, by adding three-loop evp, and by extending to an arbitrary nuclear spin. 
		Finally, we note that rotational states of antiprotonic atoms can be used to determine the mean square 
		nuclear charge radius much more accurately than from electronic or muonic atoms.
	\end{abstract}
	\maketitle
	
	\section{Introduction}
	
    Two-body systems like hydrogen \cite{hydrogen}, muonic hydrogen \cite{muonic}, or muonium \cite{muonium}
    are the most viable candidates for tests of the Standard Model of fundamental interactions at low energies. 
    These simple systems allow for highly accurate theoretical predictions. Therefore, from the comparison of theory with precise experimental data
    we are able to search for new physics and to determine values of fundamental constants such as the electron mass, 
    the nuclear magnetic moments, and nuclear charge radii \cite{codata22}. 
    
    From the theoretical side, the Schr\"odinger equation can be solved exactly for  two-body  systems and, 
    within the framework of Nonrelativistic QED (NRQED), the relativistic and QED corrections can be accounted for perturbatively
    using the exact form of the wave function. The main limiting factor for theoretical predictions comes from the not well known
    nuclear structure effects, with the exception of muonium, where the limiting factor is the electron-muon mass ratio.
    
    In atoms where the orbiting electron is replaced by a heavier particle -- such as muonic hydrogen -- vacuum polarization constitutes the most significant QED contribution to 
    energy levels. The conventional approach treats this effect perturbatively \cite{muonic}.
    In the present work, we pursue an alternative strategy: following \cite{Borie_old} we include the vacuum polarization contributions 
    directly into the Schr\"odinger equation and solve it numerically.
    The numerically obtained wave function is then used for the perturbative
    calculation of the relativistic correction, for which an exact formula
    that includes vacuum polarization potentials is known \cite{Veitia}.
    The higher-order corrections are obtained without evp, using analytically derived formulas \cite{Zatorski:22}
    which are valid for an arbitrary masses of constituent particles.
    
    Our main focus is on antiprotonic atoms -- exotic systems in which an antiproton orbits the nucleus. 
    These atoms are promising candidates  \cite{PAX, hadronic} for exploring possible new physics,
    such as long-range interactions between hadrons. Apart from this, QED effects are significantly enhanced in antiprotonic atoms 
    with a high nuclear charge $Z$, offering opportunities to verify various QED theoretical approaches through comparison with
    upcoming antiprotonic measurements planned by PAX collaboration. 
    
    In two-body systems that consist of two hadronic particles, strong interactions between the constituent hadrons must, in principle, also be considered. 
    To avoid this, we focus only on excited circular states where such an interaction, along with the nuclear structure effects, is highly suppressed.
    Surprisingly, due to the presence of evp, the mean square charge radius can contribute --as large as $10^{-5}$-- to transition energies
    between circular states. Moreover, the nuclear polarizability shift, which comes mainly from the static electric dipole polarizability $\alpha_E$, 
    contributes only at the order of $(Z\,\alpha)^6$ and  can be accounted for. This means that nuclear mean square charge radii
    can be accurately determined from circular transitions in antiprotonic atoms, provided measured values are accurate enough, typically $10^{-8}$ relative precision.
    This is slightly beyond the planned accuracy of the PAX collaboration, so further developments in precision spectroscopy of X-rays would be needed.
    
   This work is organized as follows. In Section \ref{II} we present a brief overview of the NRQED
    approach and the vacuum polarization contributions taken into account for solution of
    the Schr\"odinger equation. Section \ref{num} is dedicated to the method of numerical solution for nonrelativistic energy and wave function. 
    In Section \ref{res} we present particular results for several two-body antiprotonic systems. 
    Further, in the Supplemental material \cite{supplement} we provide the {\sl Mathematica} code {\sl PbarSpectr} that can be used to obtain theoretical predictions for energy levels
    of arbitrary excited states of antiprotonic systems with a spinless nucleus.

	\section{NRQED expansion for energy levels} \label{II}
	In the framework of NRQED the energy levels of light atoms can be expressed as a series in powers of the fine structure constant $\alpha$,
	\begin{equation}
		E(\alpha) = E^{(2)} + E^{(4)} + E^{(5)} + E^{(6)} + E^{(7)}+ E^{(8)}+ \ldots,
	\end{equation}
	where each contribution $E^{(j)}$ is of the order $\alpha^j$ and 
	may depend also on $\ln\alpha$. $E^{(2)} = E$ is the
	 nonrelativistic energy obtained by solving the Schr\"odinger equation.
	 Each higher-order term $E^{(j)}$ is calculated as an expectation value of an effective Hamiltonian $H^{(j)}$ with nonrelativistic wave function $\phi$,
	\begin{equation}
		E^{(j)} = \langle\phi|H^{(j)}|\phi\rangle + \ldots,
	\end{equation}
	where the ellipsis stands for possible second-order contributions.
	
	For a spherically symmetric potential the two-body nonrelativistic wave function $\phi(r,\theta,\phi)$ can be decomposed into a product 
	of the radial function $R(r) = P(r)/r$ and the spherical harmonic function $Y_{lm}(\theta,\phi)$. The radial part satisfies the one-dimensional  equation
	\begin{equation}\label{01}
		P''(r) + 2\,\bigg(\mu\,\big(E-V(r)\big)-\frac{l(l+1)}{2\,r^2}\bigg)\,P(r) = 0\,,
	\end{equation}
	where $\mu$ is the reduced mass and we assume natural units  $\hbar=c=1$.
	Potential $V(r)$ consists of a Coulomb potential $V_C(r) = -Z\,\alpha/r$ and the electron vacuum polarization correction $V_{\rm VP}(r)$.
	
	The electron vacuum polarization modifies the photon propagator
	\begin{equation}
	-\frac{g^{\mu\nu}}{k^2} \rightarrow -\frac{g^{\mu\nu}}{k^2\,[1+\bar{\omega}(k^2/m_e^2)]}\,,
	\end{equation}
where $k^2 = (k^0)^2 - \vec k^2$ is the photon momentum squared and $m_e$ is the electron mass.
The sum of one-particle irreducible diagrams $\bar\omega$ is expanded in a power series in $\alpha/\pi$,
\begin{equation}
\bar\omega = \bar\omega^{(1)} + \bar\omega^{(2)} + \bar\omega^{(3)} + \ldots\,,
\end{equation}
and the photon propagator is thus expanded as
\begin{equation}
-\frac{g^{\mu\nu}}{k^2} \rightarrow-\frac{g^{\mu\nu}}{k^2} \big(1 + \rho^{(1)} + \rho^{(2)} + \rho^{(3)} + \ldots)\,,
\end{equation}
with individual contributions $\rho^{(i)}$ being
\begin{align}
\rho^{(1)} = &\ -\bar\omega^{(1)}, \\
\rho^{(2)} = &\ -\bar\omega^{(2)} + (\bar\omega^{(1)})^2, \label{08} \\
\rho^{(3)} = &\ -\bar\omega^{(3)} + 2\,\bar\omega^{(1)}\,\bar\omega^{(2)} - (\bar\omega^{(1)})^3.
\end{align}
Each $\rho^{(i)}$ generates electronic vacuum polarization potential $V^{(i)}(r)$ at $k^0 = 0$,
\begin{equation}
V^{(i)}(r) = - Z\,\alpha\int\frac{d^3k}{(2\,\pi)^3}\,\frac{4\pi}{\vec k^2}\,\rho^{(i)}(-\vec k^2)\,e^{i\,\vec k\cdot\vec r }\,.
\end{equation}
For example, the one-loop vacuum polarization correction is given by the Uehling potential
	\begin{align}
		V^{(1)} = &\ -\frac{Z\,\alpha}{r}\frac{\alpha}{\pi}\int_4^\infty \frac{d(\xi)^2}{\xi^2}\,e^{-m_e\xi\,r}\,u(\xi^2),
	\end{align}
	where
	\begin{equation}
	u(\xi^2) = \frac13\sqrt{1-\frac{4}{\xi^2}}\bigg(1+\frac{2}{\xi^2}\bigg)\,,
	\end{equation}
	and $u$ is related to $\bar\omega^{(1)}$ by
	\begin{equation}
	\bar\omega^{(1)}(\zeta^2) = \frac{\alpha}{\pi}\,\zeta^2\int_4^\infty d(\xi^2)\frac{1}{\xi^2(\xi^2-\zeta^2)}u(\xi^2)\,.
	\end{equation}
	It is possible to express the Uehling potential in a closed form \cite{Frolov:24}
	\begin{align}\label{uehling}
		V^{(1)} = &\ -\frac{Z\,\alpha}{r}\frac{2\,\alpha}{3\,\pi}\bigg[\bigg(1+ \frac{b^2\,r^2}{3}\bigg)\,{\rm K}_0(2\,b\,r)
		\nonumber \\ &\
		- \frac{b\,r}{6}\,{\rm Ki}_1(2\,b\,r)
		-\bigg(\frac56 + \frac{b^2\,r^2}{3}\bigg)\,{\rm Ki}_2(2\,b\,r)\bigg]\,,
	\end{align}
	where $b=m_e/(\mu\,\alpha)$,  K$_0(z)$ is a modified Bessel function of the second kind, and Ki$_n(x)$ are Bickley functions,
	\begin{equation}
	{\rm Ki}_0(z) = {\rm K}_0(z),\,\,\,{\rm Ki}_n(z) = \int_z^\infty {\rm Ki}_{n-1}(x)\,dx\,.
	\end{equation}
	On a two-loop level, the potential
	$V^{(2)}(r)$ stands for a sum of irreducible and reducible parts of the K\"all\'en-Sabry potential \cite{KS}, see Eq. (\ref{08}). 
	Furthermore, we also include the Wichmann-Kroll correction $V_{\rm WK}(r)$ \cite{WK},
	so the complete vacuum-polarization potential is

	\begin{equation}
	V_{\rm VP}(r) = V^{(1)}(r) + V^{(2)}(r) + V_{\rm WK}(r) + \ldots,
	\end{equation}
	where the ellipsis stands for the omitted three-loop  vacuum polarization correction.
	The corresponding  uncertainty is estimated as
	\begin{equation}\label{3lest}
	\delta E_\mathrm{3loop} \approx \bigg(\frac{\alpha}{\pi}\bigg)^2 \bigg(E+\frac{(Z\,\alpha)^2}{2\,n^2}\bigg)\,.
	\end{equation}
	The exact analytic formula for $u^{(3)}$ has recently been derived \cite{onishschenko}, so it will be included in the next edition of {\sl PbarSpectr}.

	We will solve Eq.~(\ref{01}) using numerical methods described in Section \ref{num} and Appendix \ref{app:a} 
	to obtain nonrelativistic energy $E$ and the radial wave function $P(r)$.
	This function is next used for evaluation of the relativistic correction
	\begin{equation}
	E^{(4)} = \langle H^{(4)}\rangle\,,
	\end{equation}
	given by the expectation value of Breit Hamiltonian $H^{(4)}$. 
	Although we will omit the nuclear spin in the following, we first present the complete Breit Hamiltonian 
	that includes all possible interactions for spin $0$ and $1/2$ nuclei.
	For an orbiting particle with spin $s_1$, mass $m$, finite charge radius 
	$r_{C1}$ and $g$-factor $g_1$, and nucleus with spin $s_2$, mass $M$,
	nuclear charge radius $r_{C2}$ and $g$-factor $g_2$, the Breit Hamiltonian is \cite{Veitia}
\begin{widetext}
	\begin{align}
	H^{(4)} = &\ - \frac{p^4}{8}\bigg(\frac{1}{m^3}+\frac{1}{M^3}\bigg)
	+\frac18\bigg(\frac{\delta_{s_1}}{m^2}+\frac{\delta_{s_2}}{M^2} + \frac{4}{3}\,(r_{C1}^2+r_{C2}^2)\bigg)\nabla^2 V
	+\bigg[\bigg(\frac{g_1-1}{2m^2} + \frac{g_1}{2m M}\bigg)\vec L\cdot\vec s_1 \nonumber \\
	&\ \bigg(\frac{g_2-1}{2M^2} + \frac{g_2}{2m M}\bigg) \vec L\cdot\vec s_2 \bigg] \frac{V'}{r} 
	+ \frac{1}{2mM}\bigg[\nabla^2\bigg(V - \frac14 (r V)'\bigg) + \frac{V'}{r}\vec L^2 + \bigg\{\frac{p^2}{2}\,,\,V-r V' \bigg\}\bigg] \nonumber \\
	&\ + \frac{g_1\,g_2}{6mM}\vec s_1\cdot\vec s_2\,\nabla^2 V
	-\frac{g_1\,g_2}{2(2l-1)(2l+3)mM}s_1^i s_2^j\big(L^i\,L^j\big)^{(2)}\,\bigg(\frac{V'}{r}-V''\bigg)\,.
	\end{align}
	\end{widetext}
	Here, the symbol $V'$ stands for the derivative of the potential $V$ with respect to the radial distance $r$, 
	\begin{equation}
	(L^i\,L^j)^{(2)} = \frac12\{L^i,L^j\} - \frac{\delta^{ij}}{3}\,\vec L^2 
	\end{equation}
	is a symmetric and traceless irreducible tensor of the second rank,
	and $\delta_{s}=1$ for a spin $s=1/2$ particle and $\delta_{s}=0$ for a  spinless $s=0$ particle. 
	The inclusion of the evp potential into the Breit-Pauli Hamiltonian was first performed by Borie in Ref. \cite{Borie_old},
	but her result was not fully correct. The first correct derivation was performed by one of us in Ref. \cite{muonicH}
	by construction of the Breit-Pauli Hamiltonian for a massive photon. 
	This formula has been verified in many later works, including very recent ones \cite{adkins_evp},
	(cf Eq.~(21) of Ref.~\cite{muonicH} with Eqs.~(63a)-(63f) of Ref.~\cite{adkins_evp}).
	However, this  Breit-Pauli Hamiltonian for a massive photon was not suitable for a direct evaluation with the numerical wave function
	since it needs to be integrated over the photon mass, as $V^{(1)}$ in Eq. (11).
	Interestingly, the integration over the photon mass can be performed on the Hamiltonian level, see Ref. \cite{Veitia}, leading to  Eq. (19).
	Despite the correctness of the original formulas in Ref. \cite{muonicH}, Adkins and Jentschura in Ref. \cite{adkins_evp} 
	claim that the work \cite{Veitia} ``apparently omitted contributing terms due to the gauge used", which we find very awkward.

	In the following we will restrict ourselves to spinless nuclei and spin-1/2 orbiting particles.
	When calculating the expectation value of this Hamiltonian, it is advantageous to use the Schr\"odinger 
	equation and move derivatives from the potential into the wave function.
	Thus, we use the expectation value identities
    \begin{align}
    	\langle p^4\rangle = &\ 4\,\mu^2\,\langle(E-V)^2 \rangle\,, \\
    	\langle \nabla^2 V\rangle = &\ \langle -4\,\mu\,(E-V)\,V + 2 \,\vec p ~V \,\vec p ~\rangle\,, \\
    	\langle\nabla^2 (r\,V)' \rangle = &\ \langle -4\,\mu\,(E-V)\,(r\,V)'+ 2 \,\vec p\,(r\,V)'\,\vec p~\rangle\,.
    \end{align}
    With help of integration by parts we get, for instance,
    \begin{align}
    	\bigg\langle\frac{V'}{r}\bigg\rangle 
	= &\  \int dr\,P^2\,\frac{V'}{r} \nonumber \\ 
    	= &\ -\int dr\,\bigg(\frac{P^2}{r}\bigg)'\,V.
    \end{align}
    Terms with product $V\,V'$ can be handled with the help of the trivial observation that $V\,V' = \frac12\,(V^2)'$.
		Similarly, it can be shown that
	\begin{align}
	\langle \vec p\,\,f(r)\,\vec p\,\rangle = &\ \int dr\,\bigg[  \biggl(P'-\frac{P}{r}\biggr)^2+ \frac{l\,(l+1)}{r^2}\,P^2(r)\bigg]\,f(r),
\\
	\langle \vec p\,\,f'(r)\,\vec p\,\rangle =&\ 
	2\,\int dr\,\biggl(P'-\frac{P}{r}\biggr)\,\biggl[P'-\frac{P}{r} 
	\nonumber \\ &\ 
	+2\,r\,P\,\biggl(\mu\,(E-V)-\frac{l\,(l+1)}{r^2}\biggr) \biggr]\,\frac{f(r)}{r}.
	\end{align}
	In this way we move all derivatives with respect to $r$ from the potential $V$ into the wave function. 
	Any higher-order derivatives of $P(r)$ are resolved with help of the Schr\"odinger equation (\ref{01}), so that all the expectation 
	values can be expressed using only $P(r)$, $P'(r)$, and $V(r)$.
	
	For higher-order contributions of order $\alpha^{5+}$ we use results valid for two-body systems without evp.
	$E^{(5)}$ is in the case of spinless nucleus and angular momentum $l>0$ given by
	\begin{align}
		E^{(5)} = &\-\frac{7\,(Z\,\alpha)^5}{3\,\pi\,\,m\,M}\,\frac{\mu^3}{l(l+1)(2l+1)n^3}
		\nonumber \\ &\
		-\frac{4\,\alpha(Z\alpha)^5\,\mu^3}{3\,\pi\,n^3}\,\biggl(\frac{1}{m}+\frac{Z}{M}\biggr)^2\,\ln k_0(n,l)\,, 
	\end{align}
	where $\ln k_0(n,l)$ is the Bethe logarithm. 
	For $n\leq 20$ and $l\leq 19$ we take the values for Bethe
	logarithm from \cite{drake:bethelog} while for $n>20$ states
	we calculated them by ourselves.
	In the case of antiprotonic atoms, we are interested in highly excited Rydberg states for which 
	the effects of strong interaction between the nucleus and the orbiting particle are negligible. 
	We thus focus only on states with $l>1$ where the contact interaction terms vanish. 
	The next-order contribution is for $l>1$ states and spinless nucleus given by \cite{Zatorski:22}
	\begin{widetext}
\begin{align}
	E^{(6)} =&\ \mu\,(Z \alpha)^6 \biggl[\frac{X_3}{(2\kappa+1)^3\,|\kappa|\,n^3} + \frac{X_4}{(2\kappa+1)^2\,\kappa^2\,\,n^4} + \frac{X_5}{(2\kappa+1)\,|\kappa|\,n^5} + \frac{X_6}{n^6}
	\nonumber \\ &\ \hspace*{9ex}
	+ \frac{2\,\mu^3\,\big(\alpha_{E1}+\alpha_{E2}\big)}{(2l-1)(2l+1)(2l+3)}\bigg(\frac{1}{n^5}-\frac{3}{l(l+1)n^3}\bigg)\bigg]\,, \label{results0s12}
\end{align}
	where $e^2\,\alpha_{E1}$, $(Z\,e)^2\,\alpha_{E2}$ is the electric dipole polarizability of the antiproton and of the nucleus, correspondingly,
	and  $X_i$ are coefficients for a two-body system with a point spin $0$ and spin $1/2$ particles with an arbitrary g-factor
\begin{align}
X_6 =&\ -\frac{5}{16} + \frac{3}{16}\,\frac{\mu^2}{m\, M} - \frac{1}{16}\,\frac{\mu^4}{(m\,M)^2}\,,
\\
X_5 =&\ \frac{g_1^2}{(2\kappa-1)(2\kappa+3)}\,\bigg[\frac{\mu^2}{m^2}\,\frac{(3+\kappa)}{8}
+\frac{\mu^3}{m^3}\,\frac{\kappa}{4}\bigg]
+g_1\,\bigg[\frac{\mu}{m}	+\frac{\mu^2}{m^2}\,\frac{3\,(1-2\kappa-2\kappa^2)}{2\,(2\kappa-1)(2\kappa+3)}
+\frac{\mu^3}{m^3}\,\frac{3\,(-3+3\kappa+4\kappa^2)}{4\,(2\kappa-1)(2\kappa+3)}\bigg]
\nonumber \\ &\
 + \frac{3\,\kappa}{2}
+ \frac{\mu}{m}\,\frac{\kappa\,(3-8\kappa-8\kappa^2)}{2\,(2\kappa-1)(2\kappa+3)}
+ \frac{\mu^2}{m^2}\,\frac{(15-26\kappa-4\kappa^2+16\kappa^3)}{4\,(2\kappa-1)(2\kappa+3)}
+\frac{\mu^4}{m^3\,M}\,\frac{(-9+10\kappa+12\kappa^2)}{4\,(2\kappa-1)(2\kappa+3)} \,,\\
X_4 =&\ -\frac{3}{8}\,\biggl( 2\kappa + g_1\,\frac{\mu}{m} - \frac{\mu^2}{m^2}\biggr)^2 \,, \\
X_3 = &\ \frac{g_1^2}{(1+\kappa)(2\kappa-1)(2\kappa+3)} \Bigg[
- \frac{\mu^2}{m^2}\,\frac{\big(-3 - 5\kappa + 49\kappa^2 + 96\kappa^3 + 36\kappa^4\big)}{8\,\kappa^2}
- \frac{\mu^3}{m^3}\,\frac{3\,(2\kappa+1)^2}{4} \Bigg] 
\nonumber\\ &
+ g_1\, \Bigg[ \frac{\mu^2}{m^2}\,\frac{3\,(2\kappa+1)^2}{(2\kappa-1)(2\kappa+3)}
+ \frac{\mu^3}{m^3}\,\frac{\big(-3 - 5\kappa + 43\kappa^2 + 60\kappa^3 - 36\kappa^4 - 48\kappa^5\big)}
{4\,\kappa^2 (1+\kappa)(2\kappa-1)(2\kappa+3)}
- \frac{\mu}{m}\,\frac{(1+6\kappa+6\kappa^2)}{2\,\kappa(1+\kappa)}
\Bigg]\nonumber \\
&\
- \kappa + \frac{\mu}{m}\,\frac{6\,\kappa\,(2\kappa+1)^2}{(2\kappa-1)(2\kappa+3)}
+ \frac{\mu^3}{m^3}\,\frac{3\,(2\kappa+1)^2}{2\,(1+\kappa)(2\kappa-1)(2\kappa+3)}
+ \frac{\mu^4}{m^4}\,\frac{(3+2\kappa - 48\kappa^2 - 24\kappa^3 + 48\kappa^4)}
{8\,\kappa^2(2\kappa-1)(2\kappa+3)}
\nonumber\\ &\ 
- \frac{\mu^2}{m^2}\,\frac{(3+14\kappa+2\kappa^2+12\kappa^3+72\kappa^4+48\kappa^5)}
{2\,\kappa\,(1+\kappa)(2\kappa-1)(2\kappa+3)}\,,
	\end{align}
		\end{widetext}
	where 
	\begin{equation}
		\kappa = (l-j)(2j+1) = \left\{
		\begin{array}{ll}
		l & \mbox{\rm for}\; j=l-\frac{1}{2}\\
		-l-1 & \mbox{\rm for}\; j=l+\frac{1}{2}
		\end{array}\right.
	\end{equation}
	We note that  Eq.~(\ref{results0s12}) is valid for an arbitrary mass ratio $m/M$.

	The next term of the order $\alpha^7$ can be estimated by the logarithmic contribution taken from 
	the nonrecoil (infinitely heavy nucleus) hydrogenic results \cite{Jentschura:05,patkos:24} with $g=2$,
	\begin{equation}
	E^{(7)} = \frac{m\,\alpha\,(Z\alpha)^6}{\pi}\frac{8\,\big(3n^2-l(l+1)\big)\,\ln\big[\frac12(Z\alpha)^{-2}\big]}{3\,n^5\,l\,(l-1)\,(2l-1)\,(2l+1)\,(2l+3)}\,,
	\end{equation}
   	 and we assume $50\%$ uncertainty.
	For atoms with a large nuclear charge $Z$ it is also necessary to include the contribution
	of the order $\alpha^8$, which may be even more significant than $E^{(7)}$.
	We estimate it by expanding  the nonrecoil result from the Dirac equation,
	\begin{equation}
	E_D = m \bigg[1+\frac{(Z\alpha)^2}{(n-\delta)^2}\bigg]^{-1/2},
	\end{equation}
	where $\delta = j+\frac12 - \big[\big(j+\frac12\big)^2 - (Z\alpha)^2 \big]^{1/2}$, 
	up to the order $(Z\alpha)^8$
	\begin{align}
	E^{(8)} =&\  \frac{m\,(Z\alpha)^8}{16}\,\bigg(\frac{35}{8\,n^8} - \frac{15}{|\kappa|\,n^7}
	+\frac{15}{\kappa^2\,n^6} - \frac{1}{|\kappa|^3\,n^5} \nonumber \\
	&\ -\frac{3}{\kappa^4\,n^4} - \frac{1}{|\kappa|^5\,n^3} \bigg)\,,
	\end{align}
       and we assume $50\%$ uncertainty as for $E^{(7)}$.

	\section{Numerical method} \label{num}
	
	To obtain an accurate numerical solution of the Schr\"odinger equation, we have to account for the logarithmic singularity of the evp potential.
	The small $r$ expansion of the Uehling potential $V^{(1)}(r)$ is of the form
	\begin{align}
	V^{(1)}(r) =&\ \frac{Z\,\alpha^2}{\pi}\,\biggl( \sum_{i=-1}^\infty A_{2\,i+1}\,r^{2\,i+1} + A_0 + A_2\,r^2 \biggr) 
	\nonumber \\ &\ 
	+ \frac{Z\,\alpha}{\pi}\,[\gamma+\ln(m_e\,r)]
	\sum_{i=-1}^\infty B_{2\,i+1}\,r^{2\,i+1}\,.
	\end{align}
	Expansion coefficients can be obtained from Eq.~(\ref{uehling}), and the first expansion terms are
	\begin{align}
		A_{-1} =&\ \frac59\,,
		\\
		B_{-1} =&\ \frac23\,.
	\end{align}
	Similar small $r$ expansions can be obtained also for two-loop vacuum polarization \cite{KS,Martinez:11} and Wichmann-Kroll \cite{Nekipelov:12} corrections.
	We can now search for the solution of the wave function $P(r)$ near the origin in the form
	\begin{equation} \label{P1}
		P(r) =  r^{l+1}\,\sum_{i=0}^\infty\sum_{j=0}^i \,a_{i,j}\,r^i\,\ln^j (m_e\,r)\,.
	\end{equation}
	Substituting this solution along with low-$r$ expansion of vacuum polarization
	potentials into the Schr\"odinger equation,  Eq.~(\ref{01}), will lead to
	recurrence relations for coefficients $a_{i,j}$ which is solved assuming $a_{0,0}=1$.
	
	For large $r$ the one-loop vacuum polarization potential $V^{(1)}\approx r^{-\frac52}\,e^{-2\,m_e\,r} $ decreases exponentially,  similarly  $V^{(2)}$,
	so for a sufficiently large distance they can be neglected. Namely, if the ratio to the Coulomb potential  is smaller than some specified threshold,
	 $10^{-20}$ for example, then evp potentials are neglected.
	Although $V_{\rm WK}$ has an $r^{-5}$ tail, it is also neglected at large $r$ by a similar procedure.
	
	The wave function for $r\rightarrow\infty$ is searched for in the form
	\begin{equation}\label{P2}
		P(r) = r^\sigma\,e^{-\lambda r} \sum_{k=0}^\infty \frac{a_k}{r^k}\,,
	\end{equation}
	with $\sigma = Z/\lambda$ and $\lambda = \sqrt{-2E}$. We will set $a_0=1$, and the remaining
	coefficients of the expansion are obtained recurrently by inserting $P(r)$ into Eq.~(\ref{01}) with the Coulomb potential only.
	
	To numerically solve the Schr\"odinger equation for all $r$, we follow the approach of \cite{Johnson} and present the details
	of the numerical procedure in Appendix \ref{app:a}.
	For the finest grid we assumed the following parameters (in atomic units): $r_0 = 3\times10^{-5}$ and $h=3\times 10^{-4}$, $N=75~000$,
	from which we proceed to obtain $r[n_\infty]$ and $r[n_c]$.
	The convergence of the algorithm is very fast and usually three iterations are sufficient
	to obtain the nonrelativistic energy for low $Z$ with 20-digit accuracy. 
	We performed the calculation using Wolfram Mathematica,  and our code {\it PbarSpectr} is included in the Supplemental material \cite{supplement}.

	\section{Results} \label{res}
	
	In Table \ref{tab:PAX} we present the theoretical results for antiprotonic transitions considered by PAX collaboration \cite{PAX}.
	The potential $V(r)$ in the Schr\"odinger equation (\ref{01}) includes the Coulomb potential,
	the Uehling potential in the form given by Eq.~(\ref{uehling}), the two-loop vacuum polarization
	potential in the form presented in Ref.  \cite{Martinez:11}, and the Wichmann-Kroll potential in the form taken
	from \cite{Nekipelov:12}. We omitted three-loop vacuum polarization, which has recently been derived  \cite{onishschenko,adkins:25}
	but is quite complicated. It is estimated by Eq.~(\ref{3lest}),
	which is identified as an uncertainty of nonrelativistic energy and represents the dominant part of the uncertainty for current theoretical predictions.
	
	It is remarkable that even for very heavy nuclei
	the NRQED expansion converges well for circular states, and we are able to reach meV precision.
	This is because for circular states the higher-order QED contributions converge as $(Z\alpha/n)^j$, which can be  small, even
	for large $Z$, where the nonrelativistic approximation would otherwise not be appropriate. Our results are in agreement with the
	ones listed in Ref.~\cite{PAX}, but are about two orders of magnitude more accurate. In addition, we present results for a point nucleus
	and the nuclear finite size contribution, which demonstrates the possibility for determination of the nuclear charge radii.	
		\begin{table*}[!ht]
		\caption{Theoretical predictions for various antiprotonic transitions between circular states in eV. 
			$E^{(2)}$ is nonrelativistic energy obtained from the numerical solution of the
			 Schr\"odinger equation (3) with inclusion of the one- and two-loop evp. The uncertainty of nonrelativistic
			 energy comes from the omitted three-loop evp.
			$\delta E^{(2)} = E^{(2)} + (Z\,\alpha)^2/(2n^2)$ is the difference between $E^{(2)}$ and nonrelativistic energy without evp.
			$E^{(4)}$(point $N$ and $\bar p$) is the leading relativistic correction with inclusion of evp for a point nucleus and a point antiproton.
			$E_\mathrm{fns}$ is the correction to energy coming from the finite charge radius of both the orbiting antiproton and the nucleus. 
			Values of the nuclear charge radii in femtometers were taken from \cite{radii}, and we neglect electric dipole polarizabilities.}
		\label{tab:PAX}
		\begin{tabular}{ldddd }
			 Term &  \multicolumn{1}{c}{$6h_{11/2}-5g_{9/2}({}^{20}\mathrm{Ne})$}  & \multicolumn{1}{c}{$6h_{11/2}-5g_{9/2}({}^{40}\mathrm{Ar})$} &
			 \multicolumn{1}{c}{$10m_{19/2}-9l_{17/2}({}^{132}\mathrm{Xe})$} &
			 \multicolumn{1}{c}{$12o_{23/2}-11n_{21/2}({}^{184}\mathrm{W})$} \\ \hline \hline
		    $E^{(2)}$  & 29175.998(1) & 97024.511(7) & 170492.20(2) & 180512.77(2)\\
		    $\delta E^{(2)}$ & 107.252(1) & 528.128(7) & 909.72(2) & 915.08(2) \\
		    $E^{(4)}$(point $N$ and $\bar p$) & -1.927 & -22.693 & 0.37 & 37.26\\
		    $E^{(5)}$ & -0.000~054 & -0.000~81 & -0.000~6 & -0.000~4 \\
		    $E^{(6)}$ &  0.000~239 &  0.013~09 & 0.020~4 & 0.036~8\\
		    $E^{(7)}$ & -0.000~004(1) & -0.000~10(3) & -0.000~1(1) & -0.000~1 \\
		    $E^{(8)}$ & 0.000~000 & 0.000~00 & 0.000~5(1) & 0.000~6(7)\\
		    $E_\mathrm{fns}$(fs $N$ and $\bar p$) & 0.078 & 1.156 & 3.28 & 3.64\\
		    Total &  29174.149(1) & 97002.987(7) &  170495.88(2) & 180553.71(2) \\[1ex]
		    Total (point $N$ + fs $\bar p$) & 29174.076(1) & 97001.896(7) & 170492.69(2) & 180550.15(2) \\
		    $E_\mathrm{fns}$(fs $N$) & 0.007992\,r_C^2 & 0.09279\,r_C^2 & 0.1390\,r_C^2 &  0.1235\,r_C^2\\
		    $r_C$~(fm) & 3.0055(21) & 3.4274(26) & 4.7859(48) & 5.3658(23) \\			\hline \hline
		\end{tabular}
	\end{table*}
	
	\section{Conclusions}
	
	We have presented calculations of energy levels of two-body antiprotonic systems using the NRQED approach, 
	and obtained theoretical predictions for selected transitions between rotational states. 
	Our results are exact through order $(Z\,\alpha)^6$ in NRQED expansion and are valid for an arbitrary
	antiproton to the nucleus mass ratio. Higher-order corrections of order $\alpha^7$ and $\alpha^8$ are included approximately
	and are significantly smaller than the uncertainty due to the omitted three-loop evp.
	Leading nonrelativistic and relativistic contributions to energy nonperturbatively include one- and two-loop evp.
	The omitted three-loop evp gives the dominant uncertainty in our results, and its inclusion would further increase
	the accuracy by two orders of magnitude.
	In the future, the theory for medium-$Z$ atoms can be further improved by including the complete $E^{(7)}$ contribution with the full mass dependence.
	For two-body systems, this has been achieved only for radiative corrections \cite{patkos:24}.
	
	Our results for circular states are sufficiently accurate even for highly charged ions,
	which are being considered in the upcoming measurements \cite{PAX}.
	Namely, the relative uncertainty of our numerical results is of the order of
	 $10^{-7}$ for $12o\rightarrow 11n$ transition in antiprotonic ${}^{184}$W and even better for transitions in lighter
	 atoms, as can be seen in Table \ref{tab:PAX}.
	Furthemore, the Mathematica code {\sl PbarSpectr} included in the Supplemental material \cite{supplement} 
	allows for application to a wide variety of antiprotonic atoms for states with angular momentum $l>1$.

	In Table \ref{tab:PAX} we present the summary for theory of antiprotonic transitions that can be used
	for determination of the corresponding nuclear charge radii. Currently, both theory and experiment are
	not accurate enough to extract nuclear charge radii with 
	sufficient precision. From a theoretical point of view it is
	possible to improve the accuracy of the calculations by two orders of magnitude
	through inclusion of three-loop vacuum polarization into the solution of
	the Schr\"odinger equation. This is feasible and will be  done in the
	near future. From an experimental point of view, the upcoming PAX collaboration aims for relative precision of $10^{-5}-10^{-6}$
	for antiprotonic transitions using microcalorimeter detectors.
	Such a precision is still lower than the current theoretical
	accuracy. However, if the precision of experiment would reach level of meV in
	the future, we would be able to extract nuclear charge radii with an accuracy that
	would compete with the one from muonic atoms, or from electron  scattering measurements.
		
	Beside the antiprotonic atoms, the next step for theory will be to extend the calculation to the case of muonic atoms.
	For muonic atoms we need to include also the contributions to $l = 0,1$ states.
	This is, however, significantly more complicated than the case of Rydberg
	states investigated here since many additional
	 higher-order QED contributions have to be taken into account \cite{muonic}.
	Nevertheless, for the nonrelativistic, relativistic, and leading QED contributions in muonic atoms the extension of our code is simple, and it will be done in the future.
	
	\begin{acknowledgments}
	We thank Adam Prystupiuk for participation at the beginning of the project.
	V.P. acknowledges support from Cooperatio Grant No. 113-04/113COOP.
	\end{acknowledgments}

	\appendix

	\section{Numerical solution of the Schr\"odinger equation} \label{app:a}

	To solve the Schr\"odinger equation numerically we follow the approach of W. Johnson \cite{Johnson}. We introduce the function $Q(r)$, which satisfies the equations
	\begin{align}\label{15}
		\frac{d P(r)}{dr} = &\ Q(r), \\ \label{16}
		\frac{d Q(r)}{dr} = &\ -2\bigg(E-V(r)-\frac{l(l+1)}{2r^2}\bigg) P(r)\,.
	\end{align}
	The solution will be searched for on a non-uniform grid defined as
	\begin{align}
		r[i] = &\ r_0\big(e^{t[i]}-1\big),\\
		t[i] = &\ i\,h,\,\,\,\, i =0,1,2,\ldots,N
	\end{align}
	with grid parameters $r_0$, $h$, and $N$ chosen in such a way as to obtain the desired accuracy of the result. Two equations for $P(r)$ and $Q(r)$ can be combined together to give
	\begin{equation}
		\frac{d y(t)}{dt} = f(y(t),t)\,, \label{19}
	\end{equation}
	where
	\begin{align}
		f(y(t),t) = &\ G(t)\,y(t),\\
		G(t) = &\ \begin{pmatrix}
			0 & b(t) \\ c(t) & 0
		\end{pmatrix},\,\,\,
		y(t) =  \begin{pmatrix}
			P(r(t)) \\ Q(r(t))
		\end{pmatrix}\,, \\
		b(t) = &\ \frac{dr}{dt},\,\,c(t) = -2\,\frac{dr}{dt}\bigg(E-V(r)-\frac{l(l+1)}{2r^2}\bigg) \,.
	\end{align}

	Integrating Eq.~(\ref{19}) between two neighboring points on the grid we get
	\begin{equation}
		y[n+1] = y[n] + \int_{t[n]}^{t[n+1]} \,G(t)\,y(t)\,dt\,.
	\end{equation}
	Using $k$-step Adams-Moulton interpolation, we obtain for the integral
	\begin{equation}\label{36}
		y[n+1] = y[n] + \frac{h}{D}\sum_{j=1}^{k+1} a[j]\,f[n-k+j]\,,
	\end{equation}
	where coefficients $a[j]$ and $D$ depend on parameter $k$. The sum on the right-hand
	side of this equation contains $f[n+1]$ and consequently also $y[n+1]$.
	To avoid this, we define a
	$2\times2$ matrix $M[n+1]$, corresponding inverse matrix $M^{-1}[n+1]$, and constant $\Lambda = h\,a[k+1]/D$ as
	\begin{align}
		M[n+1] = &\ \mathbf{1} - \Lambda\,G[n+1],\\
		M^{-1}[n+1] = &\ \frac{1}{1-\Lambda^2\,b[n+1]\,c[n+1]} \nonumber \\ &\ \times \begin{pmatrix}
			1 & \Lambda\,b[n+1] \\
			\Lambda\,c[n+1] & 1
		\end{pmatrix}\,.
	\end{align}
	Eq.~(\ref{36}) can then be rewritten as
	\begin{equation}\label{27}
		y[n+1] = M^{-1}[n+1] \bigg(y[n] + \frac{h}{D}\sum_{j=1}^k a[j]\,f[n-k+j]\bigg)\,,
	\end{equation}
	where the dependence on $y[n+1]$ is now only on the left-hand side.
	For the first $k$ steps on the grid we use the asymptotic solution for $P(r)$ given by series
	in Eq.~(\ref{P1}) and then proceed with help of Eq.~(\ref{27}) to obtain $y[n]$ until the
	classical turning point $r[n_c]$ where $E-V(r[n_c])=0$.
	Next, we use the asymptotic solution for large $r$ given by Eq.~(\ref{P2}) to get the first $k$ values $y[N], y[N-1],  \ldots, y[N-k]$ and then proceed to obtain $y[n]$ again
	until the turning point with help of
		\begin{equation}\label{28}
		y[n] = M^{-1}[n] \bigg(y[n+1] + \frac{h}{D}\sum_{j=1}^k a[j]\,f[n+1+k-j]\bigg)\,,
	\end{equation}
	where we have to take coefficients $b(t)$ and $c(t)$ with a minus sign.
	The wave function $P(r)$ has to be continuous at the turning point $r[n_c]$,
	so both the solution evolved from small $r$ by Eq.~(\ref{27}) and the one
	evolved from large $r$ by Eq.~(\ref{28}) have to coincide at the point $r[n_c]$.
	However, since
	we have chosen arbitrarily the expansion coefficients $a_0=1$ in Eq.~(\ref{P2}) and $a_{0,0}=1$ in Eq.~(\ref{P1}), it means that
	we have to rescale $y[n]$ for $n\geq n_c$ accordingly so that $P(r)$ is continuous at the
	turning point $r[n_c]$.

	Function $Q(r)$ (i.e. the first derivative of the radial function $P(r)$) which satisfies Eq.~(\ref{16}) will be
	continuous only if $E$ is the exact nonrelativistic energy.
	In the first iteration of the algorithm we set energy $E$ equal to $-Z^2/2n^2$
	and in the following steps it will be shifted towards the correct value.
	Firstly, we assume that vacuum polarization does not change the amount of zeros
	of the wave function given by $n_r = n-l-1$. If the amount of nodes is greater or lower,
	the energy is slightly shifted correspondingly and the algorithm for obtaining the wave function is repeated. For further correction of the energy we use
	the following. Consider functions $P_1(r)$ and $Q_1(r)$ that correspond to
	energy $E_1$, and $P_2(r)$ and $Q_2(r)$ that correspond to energy $E_2$, i.e. two
	different iterations of the algorithm. It follows from Eqs.~(\ref{15}) and (\ref{16})
	that
	\begin{equation}
		\frac{d}{dr}\big(Q_2(r)\,P_1(r) - P_2(r)\,Q_1(r)\big) = 2\,(E_1-E_2)\,P_1(r)\,P_2(r)\,.
	\end{equation}
	Integrating this equation first from zero to turning point $r[n_c]$, then from the turning point
	to infinity, and summing both results leads to
	\begin{equation}
		E_1 - E_2 = \frac{(Q_1^+ - Q_1^-)\,P_2(r[n_c]) + (Q_2^- - Q_2^+)\,P_1(r[n_c])}{2\,\int_0^\infty\,P_1\,P_2\,dr},
	\end{equation}
	where symbol $\pm$ stands for the value of function $Q(r)$ at the turning point $r[n_c]$
	obtained either using Eq.~(\ref{27}) from the left or using Eq.~(\ref{28}) from the right.
	We assume that energy $E_2$ is the next step in the iteration of the algorithm and
	we also assume that for this step the difference $Q_2^- - Q_2^+$ is zero, i.e. the discontinuity of $Q(r)$ at the turning point vanish. Then,
	since $P_1\approx P_2$,
	\begin{equation}\label{A17}
		E_2 \approx E_1 + \frac{(Q_1^- - Q_1^+)\,P_1(r[n_c])}{2\,\int_0^\infty\,P_1^2\,dr}.
	\end{equation}
	We thus calculate the difference $\delta E = E_2 - E_1$ given by Eq.~(\ref{A17}) and 
	add this to the current iteration of the energy until $\delta E$ is smaller than the desired limit.
	To that end, we have to calculate the normalization of the radial function
	\begin{equation}
		\int_0^\infty P(r)^2\,dr \approx \int_0^{r[n_\infty]} P^2(r)\,dr = \int_0^{n_\infty h}
		P^2(t)\frac{dr}{dt}dt\,.
	\end{equation}
	Here, $r[n_\infty]$ is the boundary point of the grid which we choose in such a way
	that the ratio of the wave function at this point and maximum value of the wave function on the grid is less than $10^{-12}$ \cite{Johnson}.
	The expression in the last equality is calculated on a uniform grid.
	To evaluate this integral numerically, we use the $l$ step trapezoid formula
	\begin{align}\label{33}
		\int_0^{m\, h} f(x)\,dx \approx &\ h\bigg(b_1(f[0] + f[m]) + b_2(f[1] +f[m-1]) \nonumber \\ &\
		+\ldots + b_l (f[l-1]+f[m-l+1]) + f[l] \nonumber \\ &\ +\ldots+f[m-l]\bigg)\,,
	\end{align}
		where $b_i$ are coefficients of the method. The algorithm for the numerical
		calculation thus runs in the following way: first, we use the estimate for the
		nonrelativistic energy given by the formula $E=-Z^2/2n^2$. With this energy we obtain
		the wave function and its derivative and calculate the correction $\delta E$.
		If this correction is larger than the desired accuracy, we shift $E$ by this correction and recalculate again the wave function and its derivative. With these new values
		we obtain again the correction $\delta E$ and repeat this procedure until the desired
		accuracy is achieved, or in other words, until the
		discontinuity of $Q(r) = P'(r)$ at the turning point 
		is sufficiently small.
	Once we obtain the nonrelativistic energy $E$, we normalize the wave function and proceed to calculate relativistic corrections.
	For expectation values of operators in the Breit Hamiltonian we again calculate the corresponding radial integrals using the trapezoid formula in Eq.~(\ref{33}).

\end{document}